 \documentstyle[]{article}
\newcommand\mathC{{\mkern1mu\raise2.2pt\hbox{$\scriptscriptstyle|$}
        {\mkern-7mu\rm C}}}
\newcommand\unit{{\rm 1\kern-3.2pt I}}

\newcommand\e{\epsilon}

\begin{document}
\title{ Quantum mechanical histories and the Berry phase}

\author{Charis Anastopoulos \thanks{charis@physics.umd.edu}
\\
 {\small Department of Physics, University of Maryland, College Park, MD20742,
 USA } \\ \\
Ntina Savvidou \thanks{ntina@ic.ac.uk} \\
{\small Theoretical Physics Group, Imperial College,} \\
{\small SW7 2BZ London, UK} \\
{\small and }\\
 {\small Center for Gravitational Physics and Geometry,} \\  {\small Pennsylvania State
 University,} \\ {\small University Park, PA16802, USA } }

\maketitle

\begin{abstract}
We elaborate on the distinction between geometric and dynamical phase in quantum
theory and we show that the former is intrinsically linked  to the quantum mechanical
probabilistic structure. In particular, we examine the appearance of the Berry phase
in the consistent histories scheme and establish that it is the basic building block
of the decoherence functional. These results are consequences of the novel temporal
structure of histories-based theories.
\end{abstract}

\pagebreak

%\narrowtext
\section{Introduction}
The consistent histories formulation \cite{Gri84,Omn8894,GeHa9093,Har93a}  of quantum mechanics focuses on
the temporally ordered properties of physical systems: these are known as histories. There are two important structural features of
quantum mechanical history theories. The first  is that there does not exist a probability measure in the space of all histories: there exists
interference between pairs of histories.

 The second is  their non-trivial temporal structure, that allows a differentiation
 between the kinematical and the dynamical aspects of time.
This is present in the quantum temporal logic formulation of
consistent histories \cite{I94}; this allows a description of
continuous-time histories \cite{IL94, ILSS98}, in which there
exist distinct generators of time translation according to whether
they refer to dynamical  or kinematical features of the histories
\cite{Sav99, Sav99b}.

In this paper, we establish that this distinction is mirrored in a
differentiation that is well known in standard quantum thory: the
one
 between the dynamical phase due to Hamiltonian
evolution and the geometric phase of Berry \cite{Ber}.
What is more, the geometric phase manifests itself strongly in the probabilistic structure
of histories:  it is the basic building block of the interference phase between pairs
of histories. These results are established in an elementary fashion
by the study of fine-grained, continuous-time histories; they can be  then
suitably generalised.

 Now, the appearance (and measurability) of the geometric
phase in the time evolution of quantum systems   is arguably one
of the most important structural features of quantum theory. Berry
showed that when a system undergoes a cyclic evolution, due to an
adiabatic change of parameters in the Hamiltonian, a contribution
in the phase appears, that is purely geometric. In particular, the
phase contribution does not depend on the details of the dynamics
but only on the loop that was transversed by the system in the
parameter space.

It was soon realised \cite{Sim83} that the Berry phase is the holonomy of a U(1)
connection on the parameter space. In fact it can be generalised for any kind of
unitary evolution on the Hilbert space, since it arises by a natural U(1) connection
on the projective Hilbert space.

The geometric phase is a measurable quantity, that does not formally correspond to a
self-adjoint operator. Furthermore it provides a paradigm and a motivation for
investigations of topological phenomena in quantum theory, as it highlights the
natural appearance of gauge structures in the quantum formalism.

The key point of Berry's result however, that sets the subject in the foundations of
quantum theory, is the following: the Berry phase   {\it has no analogue in  the
language of probability theory}.

A probabilistic theory for a physical system---either classical or quantum---has as
basic notions {\it observables, propositions and states} that are represented by
suitable mathematical objects. In classical probability theory observables are
functions on a  space $\Omega$, propositions correspond to {\it measurable subsets} of
$\Omega$ and states to probability distributions. In quantum theory these
probabilistic concepts are also fundamental: they are  represented by Hilbert space
objects. These are  self-adjoint operators for observables, projection operators for
propositions  and density matrices for states.

However, the standard quantum mechanical formalism  refers to
properties of the system {\it at a single moment of time}: it
assigns probabilities to possible instantaneous events and studies
the evolution of these single-time probabilities. In this context,
the phase of a Hilbert space vector is  not physically relevant,
as it does not enter the single-time probability assignment. When
this phase is ignored, quantum theory is {\em only} a
generalisation of probability theory, with main difference the
non-distributivity of the lattice of propositions or equivalently
the non-commutativity of the algebra of observables.

This is the attitude taken by approaches to quantum theory that attempt to write an
axiomatic framework without assuming {\em a priori} the existence of a Hilbert space,
for example, the $C^*$--approach, quantum logic schemes or the operational approach to
quantum theory.

The existence of the Berry phase, as a measurable quantity, shows
that the {\em single-time} probabilistic description does not
exhaust the physical content of quantum theory. The Berry phase
appears in distinction to the well-known phases of unitary
evolution that are generated by a Hamiltonian. Its nature is
purely kinematical and it is a manifestation of the non-trivial
topology of the space of pure quantum states, since it appears
naturally when we view the  Hilbert space $H$ of quantum theory as
a complex line bundle over the projective Hilbert space $PH$
\cite{Sim83,AnAh,Pag}. The Berry phase is then the holonomy of the
{\em natural} connection of this bundle (i.e. the connection
induced by the inner product).

It needs to be emphasised that this bundle structure is irrelevant
{\em to any probabilistic aspects of quantum theory}. In other
words, in the unitary time evolution of quantum theory there
appears an extra phase due to the topological structure of the
theory; it has no intuitive physical explanation and it has no
classical analogue---either in classical mechanics \footnote{The
Hannay angle \cite{Ha85} is an analogue in classical mechanics.
But this appears whenever certain degrees of freedom can be
ignored due  a symmetry, whereas the wave function is assumed to
give a complete description of the quantum system.} or in
classical probability theory.

In the single-time description of a quantum system the geometric phase is lost. Hence
it is rather difficult to understand its physical meaning in standard quantum theory.
However, the importance of geometric phase is more clear in a quantum theory that is
based on {\it histories}. A history is  defined at different moments of time, in
distinction to standard single-time quantum theory. Such a formulation is provided by
the consistent histories approach to quantum theory.

This approach was developed as a realist interpretational scheme for quantum theory
\cite{Gri84,Omn8894,GeHa9093,Har93a}. As such, it suffers from the generic problems of
such schemes, i.e., contextuality of predictions  about properties of the physical
system \cite{DoKe}. Nonetheless,  it provides a new insight in understanding the
appearance of the Berry phase in quantum theory, in a manner independent of any
particular interpretational scheme one may choose to employ.

The basic object of the histories formalism is a history, i.e., a sequence of
time-ordered propositions about properties of the physical system. It corresponds to
different possible scenaria of the system. The main feature, that distinguishes
quantum mechanical histories from the ones appearing in the theory of stochastic
processes (classical probability theory), is that the probabilities for histories {\it
do not satisfy the additivity condition}.
\begin{equation}
p(\alpha \vee  \beta) = p(\alpha) + p(\beta),
\end{equation}
where $\alpha$ and $\beta$ are mutually exclusive scenaria. This is due to the fact
that quantum theory is based on amplitudes rather than probability measures, and it
further implies the existence of interference between histories.

The corresponding information, together with the probabilities, is encoded in an
object called the {\it decoherence functional}. This object incorporates the
kinematics, the dynamics and the initial condition of the physical system.

In our effort to identify the role of the Berry phase in the histories scheme, we
arrived at a surprisingly simple result: the geometric phase {\it is the main building
block of the decoherence functional}. Hence, interference between histories is
ultimately to be attributed to the presence of the geometric phase. Moreover, we
showed that the distinction between geometric phase and the dynamical phase of
canonical quantum theory --i.e the one appearing due to Hamiltonian time evolution--
is a manifestation of the temporal structure of history theories: the existence of two
laws of time transformation each corresponding to the causal/kinematical and dynamical
notions of time \cite{Sav99,Sav99b}.

\section{The geometric phase}
The simplest way to demonstrate the origin of the Berry phase is in the context of
differential geometry. To this end, let us take the complex Hilbert space $H$ to be
finite dimensional ($H = {\bf C}^{n+1}$). The inner product $<z|w> = \bar{z}_a w^a $
gives a metric $ds^2 = d \bar{z}_a d z^a $ (where $a$ runs from $0$ to $n$ and $z$
refers to coordinates  with respect to a basis), from its real part, and a symplectic
form $\omega = d \bar{z}_a \wedge d z^a $ on $H$, from its imaginary part.

The metric induces the standard metric to the unit sphere $S^{2n+1}$ of all normalised
vectors. The unit sphere is a $U(1)$ principal bundle over the projective Hilbert
space $PH$, the space of rays; this structure is known as the Hopf bundle. An element
of $PH$ is represented by $[\psi]$, the equivalence class of all normalised vectors
that differ from the normalised vector $|\psi \rangle$ only with respect to a phase.
The metric on $S^{2n+1}$ induces a metric on $PH$, defined as
\begin{equation}
ds^2(PH) = \frac{1}{1 + \bar{w}_a w^a}d \bar{w}_a dw^a
\end{equation}
and an one-form $A =  i \bar{w}_a dw^a $. Here,  we have defined
coordinates such that  for $1 \leq a \leq n$, $w^a = z^a/z^0$.

In particular, the one-form $A$ is a $U(1)$ connection form for the Hopf bundle, and
it is called the Berry connection; its curvature is equal to the projection of the
symplectic form in $PH$, modulo $i$. It may be written in a coordinate independent way
as $A = i \langle \psi|d| \psi \rangle$.

We assume an arbitrary unitary time evolution $U(s)$ on the Hilbert space $H$, and we
take an initial vector $|\psi_0 \rangle$ at time $t=0$. The curve
\begin{equation}
U(s) | \psi_0 \rangle : = |\psi(s) \rangle
\end{equation}
projects to a curve $[\psi(s)]$ on the projective Hilbert space
$PH$.

If we further assume that $U(s)$ is such that at time $t$, $[\psi(t) ] = [\psi(0)] $,
i.e. we have a loop $\gamma$ on the projective space, then the phase that is
transversed on the U(1) fiber is equal to
\begin{equation}
 e^{ \int_0^t ds \langle \psi(s)| -\frac{d}{ds} - i H(s) | \psi(s) \rangle} :=
 e^{iS[\psi(.)]},
\end{equation}
where we wrote $H(s) = U^{-1}(s) \dot{U}(s)$ and $S$ is the action out of which the
Schr\"odinger equation is derived. The second term is a time dependent angle due to
time evolution.

However, the first term is purely geometrical; it depends only on the transversed
loop, and is equal to the holonomy of the Berry connection
\begin{equation}
 e^{i \int_{\gamma} A} =
\exp \left( -\int_{\gamma} \langle \psi|d \psi \rangle \right).
\end{equation}
Note that the Berry phase  does not change if we take different representatives
$|\psi \rangle$ for the equivalence class $[\psi]$.

The geometric phase may also be defined for open paths by exploiting the metric
structure on $PH$ \cite{SaBha88}. It allows us to form a loop from any path on the
projective Hilbert space, by joining its endpoints with a geodesic. The geometric
phase of the loop thus constructed is  defined to be equal to the geometric phase
associated to the open path. Hence if $\gamma = [\psi(.)]$ is a path on $PH$, its
associated geometric phase is proved to equal
\begin{equation}
 e^{i \theta_{g}[\gamma]} = \exp \left(  -\int_{t_i}^{t_f} dt
\langle\psi(t) | \dot{\psi}(t) \rangle \right) \langle \psi_i|\psi_f \rangle .
\end{equation}
This expression is meaningful only if the endpoints are not
orthogonal.

Hence, the Berry phase  is strongly related with geometric and
topological structures  of the Hilbert space of quantum theory.
These geometric structures are physically relevant because of
Born's probability interpretation: the single-time expectation
values for observables do not change with phase transformations of
the Hilbert space vector $| z \rangle \rightarrow e^{i \phi} |z
\rangle$.

\section{Histories}
A history is defined as a sequence of projection operators $\alpha_{t_1}, \ldots,
\alpha_{t_n}$, and it corresponds to a time-ordered sequence of propositions about the
physical system. The indices $t_1, \ldots, t_n$ {\it refer to the time a proposition
is asserted and have no dynamical meaning.} Dynamics are related to the Hamiltonian
$H$, which defines the one-parameter group of unitary operators $U(s) = e^{-iHs}$.

A natural way to represent the space of all histories is by defining a history Hilbert
space ${\cal V} := \otimes_{t_i} {\cal H}_{t_i}$, where ${\cal H}_{t_i}$ is a copy of
the standard Hilbert space, indexed by the moment of time to which it corresponds. A
history is then represented by a projection operator on ${\cal V}$. This construction
has the merit of preserving  the quantum logic structure \cite{I94} and highlighting the non-trivial temporal structure
of histories \cite{Sav99,Sav99b}. Furthermore, one
can also construct a Hilbert space ${\cal V}$ for continuous-time histories
\cite{IL94,ILSS98,Ana00} by a suitable definition of the notion of the tensor product.

Furthermore, to each history $\alpha$ we may associate the class operator $C_{\alpha}$
defined by
\begin{equation}
C_{\alpha} = U^{\dagger}(t_n) \alpha_{t_n} U(t_n) \ldots U^{\dagger}(t_1) \alpha_{t_1}
U(t_1).
\end{equation}

It is important to note that time appears in {\em two distinct places} in the
definition of the class operator $C_{\alpha}$: as the argument of the Heisenberg time
evolution and as the parameter identifying the time at which a proposition is
asserted. In what follows, we will show that this distinction is strongly related to
the distinction between geometric and dynamical phase.

The decoherence functional is defined  as a complex-valued function of pairs of
histories: i.e. a map $d: {\cal V} \times {\cal V} \rightarrow {\bf C}$. For two
histories $\alpha$ and $\alpha'$ it is  given by
\begin{equation}
d(\alpha, \alpha') = Tr \left( C_{\alpha} \rho_0 C_{\alpha'}^{\dagger} \right)
\end{equation}
The standard interpretation of this object is that when $d(\alpha, \alpha') = 0$ for
$\alpha \neq \alpha'$ in an exhaustive and exclusive set of histories \footnote{ By
exhaustive we mean that at each moment of time $t_i$ $\sum_{\alpha_{t_i}} \alpha_{t_i}
= 1 $ and by exclusive that $\alpha{t_i}  \beta_{t_i} = \delta_{\alpha \beta}$. Note
that by $\alpha$ we denote both the proposition and the corresponding projector.},
then one may assign a probability distribution to this set as $p( \alpha) = d(\alpha,
\alpha)$. The value of $d(\alpha, \beta)$ is, therefore, a measure of the degree of
interference between the  histories $\alpha$ and $\beta$.

\section{The geometric phase for histories}
We now consider a time interval $[t_0,t_f]$ and a history with $n+1$ time steps
$\alpha_{t_0}, \alpha_{t_1}, \ldots \alpha_{t_f}$. We assume that the projectors are
fine-grained, which means that they correspond to elements of the projective Hilbert
space
\begin{equation}
\alpha_{t_i} = | \psi_{t_i} \rangle \langle \psi_{t_i} |.
\end{equation}

We first set the Hamiltonian equal to zero. The trace of the class operator
$C_{\alpha}$ equals
\begin{equation}
Tr C_{\alpha} =  \langle \psi_{t_0} | \psi_{t_n}\rangle  \langle \psi_{t_1}|\psi_{t_0}
 \rangle \langle \psi_{t_2}|\psi_{t_1} \rangle \ldots \langle \psi_{t_{n}}|
 \psi_{t_{n-1}} \rangle
\end{equation}
and it is non-zero provided there is no value of $i$, for which the vector
$|\psi_{t_i} \rangle$ is orthogonal to $|\psi_{t_{i-1}} \rangle$.

Next, we assume that $ \max |t_j - t_{j-1}| = \delta t$, and we choose the number of
time steps $n$ very large, so that $\delta t \sim O(n^{-1})$. Then $|\phi_{t_j}
\rangle$ approximates a path $[\phi(t)]$ on $PH$. Hence,
\begin{eqnarray}
\log Tr C_{\alpha} &=& \log \langle \psi_{t_0}|\psi_{t_n}\rangle + \sum_{i=1}^n \log
\langle \psi_{t_i} | \psi_{t_{i-1}} \rangle \nonumber \\
&=& \log \langle \psi_{t_0}| \psi_{t_n} \rangle + \sum_{i=1}^n \log \left( 1 -
\langle\psi_{t_i}  | \psi_{t_i} - \psi_{t_{i-1}} \rangle \right)
\end{eqnarray}
and the limit of large $n$ yields
\begin{equation}\log Tr C_{\alpha} = \log \langle\psi_{t_0}
|\psi_{t_n}\rangle - \langle \psi_{t_i} | \psi_{t_i} - \psi_{t_{i-1}} \rangle +
 O((\delta t)^2)
\end{equation}
As $\delta t \rightarrow 0$ the sum in the right-hand side converges to a Stieljes
integral $ - \int_{t_i}^{t_f} dt \langle \psi(t)|\dot{\psi}(t)\rangle $. Hence for a
continuous path we take
\begin{equation}
Tr C_{\alpha} = e^{i \theta_g[\psi(\cdot)]} \label {TrCa}
\end{equation}

Therefore, the  map  $ \alpha \rightarrow TrC_{\alpha}$ assigns to each fine-grained
``continuous-time'' history $\alpha$ its corresponding Berry phase. In fact, the paths
$ \psi(\cdot)$ need not be continuous; it suffices that the Stieljes integral is
defined.

Furthermore, one may use the above result to define the Berry phase, associated to a
general coarse-grained history. Hence, if $ \alpha = (\hat{\alpha}_{t_1}, \ldots ,
\hat{\alpha}_{t_n})$ is a history, then we may write
\begin{eqnarray}
\alpha \rightarrow Tr \left( \hat{\alpha}_{t_n} \ldots \hat{\alpha}_{t_2} \hat{\alpha}
_{t_1} \right). \label{aTrCa}
\end{eqnarray}
This defines a map from ${\cal V}$ to the complex numbers, that {\em assigns to each
history its corresponding geometric phase}. In particular, if we decompose  the
projector $\hat{\alpha}_{t_i}$ with respect to an orthonormal  basis in the subspace,
in which it projects
\begin{equation}
\hat{\alpha}_{t_i} = \sum_r |\psi^r_{t_i} \rangle \langle \psi^r_{t_i}|,
\end{equation}
we may then write the geometric phase for the coarse-grained histories as
\begin{equation}
\sum_{r_1, \ldots r_n} e^{i \theta_g[\psi_{r_1 \ldots r_n}(\cdot)}]
\end{equation}

In the continuum limit this can be written, suggestively, as a sum over all
fine-grained paths $\psi(\cdot)$ compatible with the coarse-grained history $\alpha$
\begin{equation}
\sum_{\psi(\cdot) \in \alpha} e^{i \theta_g[\psi(\cdot)]}
\label{ctsberry}
\end{equation}
In view of this linearity, the map that assigns to each history the corresponding
Berry phase can be described by a  functional on ${\cal V}$. When ${\cal V}$ with a tensor product of
single-time Hilbert spaces, this linear functional is naturally induced by the tensor product construction.

We must note here that, our definition of the geometric phase is structurally distinct
from the standard one. The latter refers to the evolution of a {\em state} under a
dynamical law. In histories formalism, the geometric phase is defined on {\em
observables}, or, more precisely, on  {\em possible} scenaria for the physical system.
There is, therefore, no need to make any assumption about the dynamics: this
definition of geometric phase makes sense even if the dynamics is non-unitary.

\section{The structure of the decoherence functional}
The standard form of the decoherence functional incorporates the histories $\alpha$ by
means of the operator $\hat{C}_{\alpha}$. This suggests an expression for the
decoherence functional that can be written in terms of the geometric phase.

To this end, let us assume two ``continuous-time'' histories, which we shall denote as
$\alpha_{\phi(.)}$ and $\alpha_{\psi(.)}$. From Eq.\ (\ref{TrCa}), and for the
decoherence functional written for vanishing Hamiltonian, we take
\begin{eqnarray}
d(\alpha_{\psi(.)}, \alpha_{\phi(.)} )&=& \nonumber \\
&&\hspace*{-3cm}=\langle \phi(t_i) | \rho_0| \psi(t_i)
\rangle\langle \psi(t_f)  | \phi(t_f) \rangle
\!\times\!\exp\!\!\left(\!-\!\!\int_{t_i}^{t_f}\!\!\!\! \!\!\! dt
\langle \psi(t)|\dot{\psi}(t) \rangle \!+
\!\int_{t_i}^{t_f}\!\!\!\!\!\!\! dt \langle
\dot{\phi}(t)|\phi(t)\rangle \right)\!\!\! \label{DFkin}
\end{eqnarray}
The two histories form a loop on $PH$, provided that their endpoints coincide. For
example, this is the case where the density matrix $\rho_0 $ is pure, and hence equal
to an one-dimensional projector that could be considered as part of the history. From
Eq.\ (\ref{aTrCa}) we conclude that, the value of the decoherence functional is the
Berry phase, associated to this loop.

When the Hamiltonian is included, we find
 \begin{eqnarray}
\hspace*{-1cm}d(\alpha_{\psi(.)}, \alpha_{\phi(.)} ) = \langle \phi(t_i) | \rho_0|
\psi(t_i) \rangle\langle \psi(t_f) | \rho_f | \phi(t_f) \rangle e^{iS[\psi(.)] - i
S^*[\phi(.)]},
\end{eqnarray}
where the action operator $S$ \cite{Sav99} is  given by the expression
\begin{eqnarray}
S[\phi(.)] = \int_{t_i}^{t_f} dt \langle  \phi(t)|i \frac{d}{dt} - H |\phi(t)\rangle
\label{action}
\end{eqnarray}
Hence the phase change on the Hopf bundle enters the decoherence functional at the
level of the most general fine-grained histories.

Let us now note the following:

First, the appearance of the action is contingent upon the
dynamics given by a Hamiltonian. One may consider more general
dynamics: they are incorporated in the decoherence functional
through the map $\hat{\alpha}_t \rightarrow \hat{\alpha}_t(t)$
that assigns to each Schr\"odinger-picture projector
$\hat{\alpha}_t$, a corresponding Heisenberg-picture one
$\hat{\alpha}_t(t)$, at time $t$. In full generality, it suffices
that the dynamics is generated by an one-parameter family of
automorphisms of the algebra of operators on the Hilbert space
${\cal H}$ (it does not even need to be an one-parameter group).
Hence, even though the expression involving the action is
suggestive and simple, it is not as fundamental and general as Eq.
(\ref{DFkin}), which expresses the decoherence functional in terms
of the Berry phase, prior to the introduction of the dynamics. One
should keep in mind that one aim of the  histories programme is to
describe physical systems that have non-trivial temporal structure
--as arising, for instance, in quantum gravity-- and are, perhaps,
not amenable to a Hamiltonian description. The equation
(\ref{DFkin}) for the decoherence functional is of sufficient
generality to persist even in such contexts.

Second, following  our earlier reasoning, it is easy to show that the fine-grained
expressions for the decoherence functional can be used to determine its values for
general coarse-grained histories. In analogy to (\ref{ctsberry}) they read
\begin{equation}
d(\alpha, \beta) = \sum_{\psi(\cdot) \in \alpha} \sum_{\phi(\cdot) \in \beta} \langle
\phi(t_i) | \rho_0| \psi(t_i) \rangle\langle \psi(t_f) | \rho_f | \phi(t_f) \rangle
\nonumber \\e^{iS[\psi(.)] - i S^*[\phi(.)]}
\end{equation}

Finally, the knowledge of  the geometric phase---for a set of
histories and of the automorphism that implements the
dynamics---is sufficient to fully reconstruct the decoherence
functional -- and hence all the probabilistic content of a theory.
The contribution of the initial state can be obtained by convex
combinations of a pure state at the initial moment of time. What
is interesting, is that at this level {\em there is no need for
our system to be described by a Hilbert space}. All that is needed
is a space of paths---on any manifold, the $U(1)$ connection from
which the functional giving the Berry phase will be constructed
and the dynamical law in the form of an automorphism of the space
of observables. This can be an important starting point for
developing  geometric procedures for   {\em quantisation} of
quantum mechanical histories.

\section{Conclusions}
From Eq.\ (\ref{TrCa}) we notice that the Berry phase arises
solely from the {\it ordering in time} of the projection
operators, as they appear in the decoherence functional. It
eventually corresponds to the kinematical part of the action Eq.\
(\ref{action}). The Hamiltonian part appears due to
Heisenberg-type time evolution of the projectors. This distinction
is a fundamental and impressive feature of history theories that
was identified in \cite{Sav99,Sav99b}. There exist two distinct
ways, in which time appears in physical theories: as a distinction
between past and future (partial ordering property of time)  and
as the parameter underlying the evolution laws (time as parameter
of change).

One of us (N.S.) has shown that these notions of time are associated to the
kinematical and dynamical part of the action functional respectively, and there exist
distinct operators that generate time translation with respect to these two
parameters. They are an irreducible part of any theory that is based on temporally
extended objects, whether classical or quantum. This distinction is manifested in the
two different ways the time parameter appears in the definition of the class operator
$C_{\alpha}$. From Eqs.\ (\ref{TrCa}--\ref{action}) we see that this is {\it
identical} to the distinction between the geometric and the dynamical phases of
standard quantum theory. In a sense, this is the only non-trivial remnant in canonical
quantum theory of the temporal structure of history theories. The reader is referred
to \cite{Sav99b} for a fuller treatment of this issue and to \cite{SavAn} for the
merits of the quantisation scheme motivated by it.

The fact that the off-diagonal elements of the decoherence functional correspond  to
the difference in Berry phase between its histories, suggests that the current
interpretation of probabilities in the consistent histories scheme is at least
incomplete. The relative geometric phase between two histories is a measurable
quantity, while the present interpretation gives physical meaning to the values of
only the {\it  diagonal elements} of the decoherence functional.

Of course, one might argue that the geometric phase is measured only by comparing
statistical measurements in  two different {\it ensembles} of systems. As such, it may
be described as any other measurement in the scheme. However, the point we make is
that, the off-diagonal elements of the decoherence functional have a clear geometric
and operational meaning. Therefore, an interpretational scheme that ignores them might
face a truncation of the physics it addresses. In addition, the Berry phase would
constitute a quantity that  cannot be explained in terms of the properties of an
individual quantum system, even though it is {\em measured} in ensembles. This is
extremely problematic for the aims of a realist interpretation of quantum theory.

Our results   highlight the presence of the complex phases in time evolution {\it at
the purely kinematical level}, as the main contributors in the non-additivity of the
probability measure for histories.

This strongly suggests that the presence of complex numbers in quantum theory is
intrinsically linked to its {\it distinct } ``probabilistic'' structure. To see this
consider the following.

First, both classical and quantum probability theory at a single moment of time are
described by an additive measure over a lattice of propositions. But when
time-evolution takes place in quantum theory, there appear complex phases that render
the probability measure non-additive (this is the essence of the interference of
histories).

Second, the pure time evolution in standard quantum theory is of a Hamiltonian type on
$PH$; the dynamical phases that are generated by the Hamiltonian, are {\it
structurally not different} from any  angle variables of classical mechanics. There is
{\it nothing inherently complex in them}, as the Schr\"odinger equation can be written
without any reference to an $i$. On the other hand the geometric phase appears due to
the bundle structure of the quantum mechanical space of rays. The bundle structure
arises in the first place, because  single-time probabilities do not depend on phase.
Hence, even in standard quantum theory there is  an indirect  relation between the
Berry phase and the probability assignment.  This  is brought fully into focus in the
histories formalism.

We explained in the introduction, that when we are restricted in a single moment of
time, the structures of quantum theory are in one-to-one analogy with the ones of
classical probability theory. From an operational perspective, quantum mechanics at a
single moment of time may be formulated without making any reference to complex
numbers; it can be stated solely in terms of real-valued observables, expectation
values and probabilities. It is only, when we study physical systems in a temporal
sense that complex numbers appear. However, their appearance cannot be attributed to
the law of time evolution.

Dynamics, in general, appears as an automorphism of the space of observables: if the
observables are defined as real-valued, they will remain real-valued when dynamically
transformed. For example, Schr\"odinger's equation does not need introduce the complex
unit; it can be equally well written in a real Hilbert space \cite{Stuc}.

Alternatively, one may substitute Schr\"odinger's equation with a
real, partial differential equation on phase space---using either
the Wigner or the coherent state transforms. Hence, while complex
numbers in quantum theory are unavoidable when we study properties
of the system at different moment of time, {\em they are not
introduced by the dynamical law}. Furthermore, it is the temporal
ordering that introduces phases, in an irreducible way, into the
decoherence functional, that it is encoded in the definition of
the  class operator $C_{\alpha}$.

In other words, the geometric phase is a genuinely complex-valued object; and it is
only the fact that we {\em measure} such an object, that {\em forces} us to accept
complex numbers as an irreducible part of quantum theory. Complex number are not a
necessary consequence of {\em any dynamical law}.

Hence, we conclude that, {\it the complex structure of quantum theory is intrinsically
linked to both its probability structure and the way the notion of succession is
encoded.} After all quantum theory is a theory of amplitudes, and the results from the
above analysis imply that {\it all physically relevant amplitudes}---contained in the
decoherence functional---are constructed from the geometric phase. As such they are
{\it geometrical in origin}.

This is an intriguing result. It is  a structural characteristic of quantum
probability  that should persist in frameworks that attempt to generalise quantum
theory in a way that the Hilbert space is not a necessary ingredient.

\section*{ Acknowledgments:} C.A. was supported from the NSF grant PHY90-00967 and N.S.
by a gift from the Jesse Phillips Foundation.

%\widetext
\end{document}